\documentstyle[preprint,aps,prl]{revtex}
\begin{document}
\draft
\title{\bf Many-body correlations in a multistep variational approach}
\author{M. Sambataro}
\address{Istituto Nazionale di Fisica Nucleare, Sezione di Catania\\
Corso Italia 57, I-95129 Catania, Italy}

\maketitle
\begin{abstract}
We discuss a multistep variational approach for the study of many-body
correlations. The approach is developed in a boson formalism (bosons
representing particle-hole excitations) and based on an iterative sequence
of diagonalizations in subspaces of the full boson space. Purpose
of these  diagonalizations is that of searching for the best 
approximation of the ground state of the system. The procedure
also leads us to define a set of excited states and, at the same time, of
operators which generate these states as a result of their action
on the ground state. We examine the cases in which these
operators carry one-particle one-hole and up to two-particle two-hole
excitations. We also explore the possibility of associating bosons
to Tamm-Dancoff excitations and of describing the spectrum
in terms of only a selected group of these. Tests within
an exactly solvable three-level model are provided.
\newline
\end{abstract}

\newpage
\section*{1. Introduction}
Developing reliable microscopic approaches for the description of
correlations in quantum many-body systems is a field of active
research in various branches of physics. A preeminent role in this
field has been traditionally played by the Random Phase Approximation
(RPA) \cite{ring}. Over the years, however, several 
attempts \cite{hara,rowe2,provi,duke,%
schuck,klein3,cata,duke2,mars3,chomaz,samba2,cata2,grasso,dinh,%
samal} have also been made
to overcome the natural limitations of this theory related, in particular,
to its lack of an internal consistency. 

In a recent publication \cite{samal}, with reference to the $\beta$-decay
physics and working in a quasi-particle formalism, we have discussed
an approach aiming at improving the quality of the standard 
Quasi-particle RPA calculations usually made in this field.
The basic point of this approach has been that of searching first for the best
approximation of the ground state. In order to do this we have started with
an initial $Ansatz$ for this state and we have tried to improve this 
approximation through a series of minimizations 
which modified the structure of the state at each step. Excited states
have, then, been constructed by acting with an RPA-like 
phonon on the ground state
and fixing the variables of this phonon via the minimization of the energy.
Tests of the procedure have proved to be quite encouraging.
In the present work, however, we present an
evolution of this method which we believe to be 
more effective and simpler to apply in realistic cases.

The approach is developed in a boson formalism. As a preliminary
step, then, a boson space will be defined where bosons identify 
particle-hole excitations and a mapping procedure will allow 
the transformation of
fermion operators onto their images in this boson space. Similarly to
the previous work \cite{samal}, the basic point of the approach will 
consist in searching first for the best approximation of the ground
state. Differently from the mentioned case, however, this will be achieved by
means of an iterative sequence of diagonalizations 
in subspaces of the full boson space. As a further and
important difference from the case of Ref. \cite{samal}, as a result
of this sequence of diagonalizations, a set of operators will also be
generated which by acting on the ground state of the system will
define a set of excited states. We will consider two cases:
the case in which these operators carry only one-particle one-hole
(1p-1h) excitations and the case in which they include up to 
two-particle two-hole (2p-2h) excitations. In both cases the comparison with
exact calculations within a schematic model will allow us to judge 
the quality of the approximations. As a schematic model we have chosen the
SU(3) model \cite{li,matsuo,takada}.

In the second part of the work, we will reformulate the fermion-boson
correspondence and identify bosons with
properly chosen $collective$ particle-hole excitations. Then,
what is more interesting, we will explore the possibility of
describing at least partially the spectrum of the system in terms of
only a selected group of these bosons.
Once again we will first search for the best approximation of the
ground state and we will then construct excited states carrying up to
1p-1h and 2p-2h excitations. Tests within the schematic model will be
provided also in this case.

The paper is organized as follows. In Sect. 2, we will describe 
the basic points of the procedure limiting ourselves to the case of
1p-1h excitations. In Sect. 3, we will provide some applications
within the SU(3) model. In Sect. 4, we will
consider an extension of the procedure to include 2p-2h excitations.
In Sect. 5, we will examine the case of bosons as collective 
particle-hole excitations. Finally, in Sect. 6, we will summarize the
results and give some conclusions. The Appendix A will be reserved to 
describe some details of the mapping procedure.

\section*{2. The procedure: 1p-1h excitations}
To simplify the notation, we will illustrate the procedure directly within 
the exactly solvable model which has been used for our tests. This model,
the so-called SU(3) model, was first discussed by
Li et al. \cite{li} and has been used more recently by
Matsuo et al. \cite{matsuo} and Takada et al. \cite{takada} to test some
approximation schemes. The model consists of three $2\Omega$-fold degenerate
single-particle shells which are occupied by $2\Omega$
particles. In the absence of interaction, then, the lowest level is
completely filled while the others are empty. This state,
the ``Hartree-Fock" (HF) state of the system, is denoted by $|0\rangle$.
A single-particle state is specified by a set
of quantum numbers ($j,m$), where $j$ stands for the shell
($j$=0,1,2) and $m$ specifies the $2\Omega$ substates within the shell.
The creation and annihilation operators of a fermion in a state ($j,m$)
are defined by $a^{\dagger}_{jm}$ and $a_{jm}$, respectively.

Let us consider the operators
\begin{equation}
K_{ij}=\sum^{2\Omega}_{m=1}a^{\dagger}_{im}a_{jm}~~~~(i,j=0,1,2).
\label{1}
\end{equation}
These operators satisfy the Lie algebra of the group SU(3)
\begin{equation}
[K_{ij},K_{kl}]=\delta_{jk}K_{il}-\delta_{il}K_{kj}.
\label{2}
\end{equation}
It is assumed that the Hamiltonian of the model 
is written in terms of the generators
$K_{ij}$ only and contains up to two-body interactions. Its form is
\cite{takada}
\begin{eqnarray}
H_F=&&\sum_{i=1,2}\epsilon (i)K_{ii}+%
      \sum_{i,j=1,2}V_x(i,j)K_{i0}K_{0j}\nonumber\\
    &&+\frac{1}{2}\sum_{i,j=1,2}V_v(i,j)(K_{i0}K_{j0}+K_{0j}K_{0i})\nonumber\\
    &&+\sum_{i,j,k=1,2}V_y(i,j,k)(K_{i0}K_{jk}+K_{kj}K_{0i}),
\label{3}
\end{eqnarray}
where the  coefficients are real and obey the symmetry conditions 
$V_x(i,j)=V_x(j,i)$ and $V_v(i,j)=V_v(j,i)$. The eigenstates of $H_F$ are
constructed by diagonalizing it in the space
\begin{equation}
F=\left\{|n_1n_2\rangle =%
\frac{1}{\sqrt{{\cal N}_{n_1n_2}}}(K_{10})^{n_1}(K_{20})^{n_2}%
|0\rangle\right\}_{0\leq n_1+n_2\leq 2\Omega},
\label{4}
\end{equation}
where ${\cal N}_{n_1n_2}$ are normalization factors.

As in Ref. \cite{samal}, we will work in a boson formalism. To begin, then, we
define the boson space
\begin{equation}
B=\left\{|n_1n_2) =\frac{1}{\sqrt{n_1!n_2!}}%
(b^{\dagger}_1)^{n_1}(b^{\dagger}_2)^{n_2}%
|0)\right\}_{0\leq n_1+n_2\leq 2\Omega},
\label{5}
\end{equation}
where the operators
$b^{\dagger}_i$ obey the standard boson commutation relations
\begin{equation}
[b_i,b^{\dagger}_j]=\delta _{ij},~~~~~[b_i,b_j]=0
\label{50}
\end{equation}
and $|0)$ is the boson vacuum. As evident from a glance at 
(\ref{4}) and (\ref{5}), a
one-to-one correspondence exists between the states of $F$ and $B$, the
boson operators $b^{\dagger}_i$ playing the role of the excitation operators
$K_{i0}$ and the boson vacuum $|0)$ replacing the HF state
$|0\rangle$. As anticipated in the Introduction, however, 
in Sect. 5 we will also 
examine a different correspondence and so a different meaning to 
attribute to these boson operators.

The mapping procedure to construct boson images of fermion operators is
the same discussed in
previous works \cite{samba2,samal} and it is based on the requirement that
corresponding matrix elements in $F$ and $B$ be equal. The procedure
is, therefore, of Marumori-type. We will give further details in
App. A. Here, we simply say that, in correspondence
with the Hamiltonian $H_F$ (\ref{3}), we introduce 
an hermitian boson Hamiltonian $H_B$ which contains
up to five-boson terms. This has therefore the general form
\begin{eqnarray}
H_B=&&\alpha +\sum_i\beta_i(b^{\dagger}_i+h.c.)+%
      \sum_{ij}\gamma_{ij}b^{\dagger}_ib_j+%
      \sum_{i\leq j}\phi_{ij}(b^{\dagger}_ib^{\dagger}_j+h.c.)\nonumber\\
    &&+\sum_{i\leq j}\sum_k\epsilon_{ijk}(b^{\dagger}_ib^{\dagger}_j%
      b_k+h.c.)+%
      \sum_{i\leq j}\sum_{k\leq l}\delta_{ijkl}%
      b^{\dagger}_ib^{\dagger}_jb_kb_l\nonumber\\
    &&+\sum_{i\leq j\leq k}\sum_l\rho_{ijkl}(%
      b^{\dagger}_ib^{\dagger}_jb^{\dagger}_kb_l+h.c.)+%
      \sum_{i\leq j\leq k}\sum_{l\leq m}\tau_{ijklm}(%
      b^{\dagger}_ib^{\dagger}_jb^{\dagger}_kb_lb_m+h.c.).
\label{51}
\end{eqnarray}

To illustrate the iterative sequence of diagonalizations 
on which our approach is based
we start by introducing an arbitrary boson state $|\Psi^{(0)}_0)$.
We consider this as a zeroth-order approximation of
the ground state and we assume
$|\Psi^{(0)}_0)=\frac{1}{\sqrt{3}}(|0)+b^{\dagger}_1|0)+b^{\dagger}_2|0))$.
Let's then consider the space
\begin{equation}
B^{(1)}\equiv\left\{|\Psi^{(0)}_0), b^{\dagger}_i|\Psi^{(0)}_0),
b_i|\Psi^{(0)}_0)\right\}_{i=1,2}
\label{52}
\end{equation}
and diagonalize $H_B$ in this space. $B^{(1)}$ is, in general, 
considerably smaller than the full boson space $B$. In our calculations,
for instance, we have assumed $2\Omega =10$ and this implies that the 
space $B$ can have up to ten-boson states while $B^{(1)}$ contains only up
to two-boson states. However, if $|\Psi^{(1)}_0)$ denotes
 the lowest eigenstate
resulting from this diagonalization, one can only expect that 
$|\Psi^{(1)}_0)$
will provide an approximation of the ground state better than (or, at worst,
equal to) $|\Psi^{(0)}_0)$. This is due to the fact that we are allowing the 
new state to have more components than $|\Psi^{(0)}_0)$ and that the 
coefficients of $|\Psi^{(1)}_0)$
are fixed to guarantee the lowest energy of the state.
We define $|\Psi^{(1)}_0)$ the first-order approximation of the ground
state.

As a next step, let's consider the space
\begin{equation}
B^{(2)}\equiv\left\{|\Psi^{(1)}_0), b^{\dagger}_i|\Psi^{(1)}_0),
b_i|\Psi^{(1)}_0)\right\}_{i=1,2}
\label{53}
\end{equation}
and diagonalize $H_B$ in this space. If $|\Psi^{(2)}_0)$ is the lowest
eigenstate resulting from this diagonalization, the above arguments lead
us to expect that also $|\Psi^{(2)}_0)$ will be better than $|\Psi^{(1)}_0)$.
We define $|\Psi^{(2)}_0)$ the second-order approximation of the ground state.
The procedure can go on as many times as one wishes. By performing a
sequence of diagonalizations in spaces whose dimensionality remains
unchanged (and much smaller than that of the full boson space) one can
construct approximations of the ground state which improve step-by-step.

It turns out to be interesting to reformulate the procedure just described as
follows. Let's define the operator
\begin{equation}
(Q^{\dagger}_0)^{(\nu )}=\sum_iX^{(\nu )}_ib^{\dagger}_i+%
\sum_iY^{(\nu )}_ib_i+Z^{(\nu )}.
\label{54}
\end{equation}
It is, then,
\begin{equation}
|\Psi^{(0)}_0)=(Q^{\dagger}_0)^{(0)}|0),
\label{55}
\end{equation}
with $X^{(0)}_i=Z^{(0)}=\frac{1}{\sqrt{3}}$ (the
coefficients $Y^{(0)}_i$ remain undetermined in this case).
Similarly, one can define an operator $(Q^{\dagger}_0)^{(1)}$ such that
\begin{equation}
|\Psi^{(1)}_0)=(Q^{\dagger}_0)^{(1)}|\Psi^{(0)}_0)
\label{56}
\end{equation}
and so on for all other approximations. 
In general, if $|\Psi^{(k)}_0)$ denotes
the $k$-th approximation of the ground state, one can write
\begin{equation}
|\Psi^{(k)}_0)=(Q^{\dagger}_0)^{(k)}|\Psi^{(k-1)}_0)=%
(Q^{\dagger}_0)^{(k)}(Q^{\dagger}_0)^{(k-1)}\cdot\cdot\cdot%
(Q^{\dagger}_0)^{(0)}|0).
\label{57}
\end{equation}
Therefore $|\Psi^{(k)}_0)$ is a product of $k+1$ operators $Q^{\dagger}$ 
of the type (\ref{54}), $k$ corresponding to the $k$ diagonalizations in the
$B^{(k)}$ subspaces plus the operator $(Q^{\dagger}_0)^{(0)}$ corresponding
to the starting $Ansatz$ $|\Psi^{(0)}_0)$. Concerning this state, some
comments are necessary to justify its use. In principle, one could have 
started with a diagonalization similar to all the other ones, namely in
a space of the type (\ref{52}) where $|\Psi^{(0)}_0)\equiv |0)$.
However, the coefficients $\beta_i$ of the boson
Hamiltonian (\ref{51}) are nothing but the matrix elements of $H_F$
between the HF state $|0\rangle$ and the 1p-1h states 
$K_{i0}|0\rangle$ (see App. A). These coefficients turn out to be zero
in our model and the same would happen in a realistic case.
In consequence of that no mixing is possible
between the states $|0)$ and $b^{\dagger}_j|0)$ and so a diagonalization
in the space $\{|0),b^{\dagger}_j|0)\}$ could generate (what indeed happens
in our model) the boson vacuum
$|0)$ as the lowest eigenstate. This would lead to a crash of the 
iterative mechanism.

Once a sufficient number of iterations has been performed the procedure
is expected to reach convergence. If this is the case, any diagonalization
beyond a given one, let's say the $k$-th one, will have to leave the results 
unmodified. This necessarily implies that the operator $(Q^{\dagger}_0)^{(k+1)}$
which will emerge from the $(k+1)$-th diagonalization will
have coefficients
\begin{equation}
X^{(k+1)}_i=Y^{(k+1)}_i=0,~~~~~Z^{(k+1)}=\pm 1.
\label{58}
\end{equation}
Convergence of the procedure therefore means convergence towards these
values of the coefficients $X, Y$ and $Z$.

As a result of the same $(k+1)$-th diagonalization, besides the operator
$(Q^{\dagger}_0)^{(k+1)}$, one will also obtain the operators 
$(Q^{\dagger}_i)^{(k+1)}$
associated to the remaining eigenstates. The number of these
eigenstates is (up to) $2N$, where $N$ is the number of the 1p-1h
excitations ($N$=2 in our model). If we call
$|gs)$ our best approximation for the ground state, i.e. 
$|gs)\equiv |\Psi^{(k)}_0)$, these further eigenstates can 
be written as
\begin{equation}
|\Psi^{(k+1)}_i)\equiv (Q^{\dagger}_i)^{(k+1)}|gs)%
~~~~~(i=1,...,2N).
\label{59}
\end{equation}
Keeping in mind that $(Q^{\dagger}_0)^{(k+1)}=\pm 1$, one has that
\begin{equation}
(gs|(Q^{\dagger}_i)^{(k+1)}|gs)=0%
~~~~~(i=1,...,2N).
\label{60}
\end{equation}
Moreover, the operators $(Q^{\dagger}_i)^{(k+1)}$ satisfy the
orthogonality conditions
\begin{equation}
(gs|(Q_i)^{(k+1)}(Q^{\dagger}_j)^{(k+1)}|gs)=\delta_{ij}%
~~~~~(i,j=1,...,2N).
\label{61}
\end{equation}
This procedure therefore leads us to define a set of operators 
$(Q^{\dagger}_i)^{(k+1)}$ whose action on the ground state gives rise 
to a set of excited states which, considering
the nature of the operators (\ref{54}), all carry excitations of the
type 1p-1h.

This way of representing the excited states shows evident similarities
with that of RPA. The operators $Q^{\dagger}$ (\ref{54}) indeed remind us
(although in a boson formalism) of the phonon operators of RPA.
However, important differences do appear between the two approaches.
While in RPA the ground state is defined as the vacuum of the operators
$Q$, this is not true in this approach. Here, a sequence of
operations of variational type leads us 
to construct both the ground state and the
set of operators $Q^{\dagger}$ which define the excited states. In RPA,
instead, the operators $Q^{\dagger}$ are first constructed by solving 
some equations (not of variational type) and an explicit expression for the
ground state can be subsequently derived. We also notice that the 
$Q^{\dagger}$'s (\ref{54}), although constructed in terms of bosons, are not
real bosons themselves since they do not obey standard commutation relations
of the type (\ref{50}). This is not the case in RPA where, at least in the
standard quasi-boson approximation, the operators $Q^{\dagger}$ are treated
as bosons. 

Before concluding this section, some comments are necessary about the
violation of the Pauli Principle which is always a risk whenever dealing
with boson transformations. Although the sequence of diagonalizations
described above
can be extended as long as one wishes and so one can in principle form
states of the type (\ref{57}) which involve any number of operators
$Q^{\dagger}$, not all the components of these states may
be  ``physical".
In other words, there could be components, the so called ``spurious"
components, which have not a counterpart 
in the fermion space $F$. In our model, for
instance, whenever acting with more than $2\Omega$ operators $Q^{\dagger}$
on the boson vacuum $|0)$ one would form states having,
among the others, components with more than $2\Omega$ operators $b^{\dagger}_j$
and these are all spurious components.

In order to properly take into account these components, one should
in principle perform the diagonalizations in spaces of the type
\begin{equation}
B^{(k)}=\left\{|\Psi^{(k-1)}_0),%
{\widehat I_B}b^{\dagger}_i|\Psi^{(k-1)}_0),%
b_i|\Psi^{(k-1)}_0)\right\}_{i=1,2},
\label{62}
\end{equation}
where we have introduced the identity operator $\widehat{I}_B$ of the boson
space B. This is rather simple to do in the model under discussion.
However, as we will see in the next sections, the rate of convergence
of the procedure in the cases examined has always been such as to make the
problem associated to the occurrence of these spurious components
unimportant in these calculations.

\section*{3. Results}
The calculations we are going to describe refer to the following
choice of the parameters:
$2\Omega$=10, $\epsilon (1)=\epsilon$,
$\epsilon (2)=1.5\epsilon$,
$V_x(i,j)=-2\chi$, $V_v(i,j)=\frac{1}{2}\chi$ and
$V_y(i,j,k)=-\frac{3}{4}\chi$ $(i,j,k,=1,2)$.
Both $\epsilon$ and $\chi$ are parameters expressed in units of energy.

In Fig. 1, the solid lines show the ground state energy (A) and the excitation 
energies of the lowest five states (B), in units of $\epsilon$, as functions
of the strength ${\chi}/{\epsilon}$. These results are obtained by diagonalizing 
$H_F$ in $F$. Dot-dashed lines show the equivalent results for $H_B$ in $B$.
The agreement between the fermion and boson spectra guarantees the very good
quality of the boson image $H_B$.

In Fig. 2, lower part, we show the ground state energies corresponding to
different orders of approximation as indicated by the numbers which label 
the dot-dashed lines. For comparison, we plot
(solid line) the energies which result from the diagonalization of $H_B$
in $B$ since this represents the best one can hope to reproduce in this
approach. The same will be done in all the next figures. As seen in Fig. 1,
however, fermion and boson energies differ very little from each other.
A clear improvement of the quality of the approximation is observed 
in correspondence with the increasing of its order.

Still in Fig. 2, upper part, dot-dashed lines show the spectrum obtained
within this approach (the spectrum is found in correspondence with the best
approximation of the ground state shown in the lower part of the figure).
For comparison we also show the energies of the two one-phonon RPA states
(dashed lines). RPA undergoes a collapse as
soon as the ground state energy starts deviating 
significantly from zero
(${\chi}/{\epsilon}\approx 0.024$). 
The same states, but within the whole range of ${\chi}/{\epsilon}$,
are obtained within our approach and they well reproduce the exact ones.
It is worth noticing that our approximated spectrum is formed by 
four excited states (with
the only exception of very small values of the strength ${\chi}/{\epsilon}$
where they can become two)  as opposed to the two states 
of RPA. Already for a strength ${\chi}/{\epsilon}\agt 0.014$ the second
one-phonon RPA state actually corresponds 
to the third excited state within the present approach. The forth
approximate state lies higher in energy and has not been reported in the
figure.

\section*{4. 2p-2h excitations}
The same procedure discussed in Sect. 2 and involving only 1p-1h excitations 
can be extended in a natural way to include 2p-2h excitations as well.
The basic difference 
consists in performing each diagonalization of the iterative
sequence in spaces of the type
\begin{equation}
B^{(k)}=\left\{|\Psi^{(k-1)}_0),%
b^{\dagger}_i|\Psi^{(k-1)}_0),%
b^{\dagger}_ib^{\dagger}_j|\Psi^{(k-1)}_0),%
b_i|\Psi^{(k-1)}_0),%
b_ib_j|\Psi^{(k-1)}_0)%
\right\}_{i\leq j=1,2}.
\label{63}
\end{equation}
Moreover, differently from the case of Sect. 2, there is no more need for
an initial $Ansatz$ for the ground state. The iterative procedure simply
 begins by
performing a diagonalization in a space $B^{(1)}$ of the form $(\ref{63})$
where $|\Psi^{(0)}_0)\equiv |0)$. 

To evaluate the role of these additional excitations in the structure of
the spectrum, we have performed calculations similar to those shown in
Fig. 2. The new results are reported in Fig. 3. For 
what concerns the ground state energy one observes a very good agreement and
a faster convergence with respect to the 1p-1h case. Concerning the
spectra of the lowest five excited states, besides the two ``one-phonon"
states discussed in the previous section, also the remaining states
are now well reproduced (as for Fig. 2, this spectrum refers to the best
approximation of the ground state as indicated in the 
lower part of the figure). As expected, then, the inclusion of 2p-2h
excitations considerably improves the quality of the approximate spectrum.

\section*{5. Bosons as collective partilcle-hole excitations}
When performing the boson mapping we have established a one-to-one
correspondence between the states $|n_1n_2\rangle$ and $|n_1n_2)$
defined in the Eqs. (\ref{4}) and (\ref{5}), respectively. In such a
correspondence, bosons $b^{\dagger}_j$ are images of the 1p-1h
operators $K_{j0}$. However, as already anticipated, this is not the
only possibility of correspondence. 
To show an alternative choice, let's proceed as in
Takada et al. \cite{takada} and let's first define
the Tamm-Dancoff (TD) phonon operator
\begin{equation}
V^{\dagger}_{\lambda}=\frac{1}{\sqrt{2\Omega}}\sum_iv^{(\lambda )}_iK_{i0}.
\label{13}
\end{equation}
The amplitudes $v^{(\lambda )}_i$ satisfy the orthogonality condition
\begin{equation}
\sum_i v^{(\lambda )}_iv^{(\lambda ')}_i=\delta_{\lambda\lambda '} 
\label{64}
\end{equation}
and are
obtained by diagonalizing $H_F$ in the basis 
$\{\frac{1}{\sqrt{2\Omega}}K_{i0}|0\rangle \}_{i=1,2}$. In terms of these TD
operators we construct the space
\begin{equation}
\left\{\overline{|n_1n_2\rangle}=%
\frac{1}{\sqrt{{\cal N}'_{n_1n_2}}}
(V^{\dagger}_{1})^{n_1}(V^{\dagger}_{2})^{n_2}%
|0\rangle\right\}_{0\leq n_1+n_2\leq 2\Omega}.
\label{14}
\end{equation}
This space is the same as (\ref{4}) but just a different representation.
Therefore, if we establish a one-to-one correspondence between states
(\ref{14}) and (\ref{5}) and we reconstruct the boson image of $H_F$,
the new boson Hamiltonian will have different coefficients 
(for instance, the matrix $\gamma_{ij}$ of (\ref{51}) will now be
forced to be diagonal)
but its spectrum will remain unchanged.
In this new representation, bosons correspond to $collective$
particle-hole excitations and so play a role very similar to that of the 
standard s,d,... bosons in the Interacting Boson Model picture \cite{iach}
(where they are meant to represent collective 
particle-particle excitations). As in this case, then, it is
natural to expect that the structure of the low-lying part of the
spectrum may be described in terms of only a selected group of collective
bosons.

Our model appears particularly suited to illustrate this point. 
One can form two TD excitations (\ref{13}) and their energies are
shown in Fig. 4. As one sees, while increasing the strength
${\chi}/{\epsilon}$, one of the energies remains almost constant while
the other one shows a regular and sizeable decrease. 
This behaviour leads us to believe
that the lowest boson may play a preeminent role in the structure
of the low-lying spectrum. 
We have made some
calculations involving only this boson and (to start) only 1p-1h ecitations. 
By denoting $b^{\dagger}(b)$
the creation(annihilation) boson operator associated to the lowest TD
excitation, the procedure consists now in an iterated sequence
of diagonalizations in spaces of the type
\begin{equation}
B^{(k)}=\left\{|\Psi^{(k-1)}_0),%
b^{\dagger}|\Psi^{(k-1)}_0),%
b|\Psi^{(k-1)}_0)\right\}.
\label{65}
\end{equation}
As initial $Ansatz$ we have assumed the state 
$|\Psi^{(0)}_0)=\frac{1}{\sqrt{2}}(|0)+b^{\dagger}|0))$. The new results
are shown in Fig. 5. The agreement for the ground state and the first
excited state is very good. It is worth stressing that this result has been
obtained by working in spaces whose dimensionality 
is remarkably smaller than that of the full boson space $B$
(3 versus 66 in our model).

In Fig. 6, we show similar calculations which involve up to 2p-2h 
excitations. The quality of the agreement for the ground state and the first
excited state (already good) remains basically unchanged while also
the lowest ``two-phonon" state (to use an RPA language) 
is now well described. These calculations therefore
confirm the expectation that only the boson associated to the lowest TD
excitation plays an active role in the structure of these states.

\section*{Summary and conclusions}

In this paper we have presented a multistep variational approach for
the study of many-body correlations. The approach has been developed
in a boson formalism (bosons representing particle-hole excitations)
and based on an iterative sequence of diagonalizations in subspaces of 
the full boson space. Purpose of these diagonalizations has
been that of searching for the best approximation
of the ground state of the system. The procedure has also led us to define
a set of excited states and, at the same time, of operators 
generating these states as a result of their action on the ground state.
We have considered two cases: a), the case in which these operators carried
only 1p-1h excitations and, b), the case in which also 2p-2h
excitations were included.

The approach has been tested within an exactly solvable three-level model.
The comparison betwen exact and approximate ground state
energies has allowed us to appreciate the convergence of the procedure
in both cases a) and b).
In the first case, a comparison also with standard RPA calculations
has shown that the ``one-phonon" states that this theory could reproduce
up to its crash-point were now well reproduced in the whole range of
variation of the strength. With the inclusion of 2p-2h excitations also
the remaining low-lying states of the spectrum have been well reproduced.

In the second part of the paper, we have reformulated the fermion-boson
correspondence and identified bosons with Tamm-Dancoff
phonons. We have then explored
the possibility of describing at least partially the spectrum of the
system in terms of only a selected group of these bosons. In our model this
has implied restricting the set of two possible bosons to the one
corresponding to the lowest excitation energy. We have verified that
the ground state was still well reproduced and so was
the first excited state already at the level of 1p-1h excitations.

The possibility of selecting a restricted set of collective particle-hole
excitations and therefore of constructing the boson space only in terms of
the corresponding bosons appears to be quite appealing. It may represent, 
in fact, an effective way to reduce the dimensionalities of the system 
and so to lead to a much
simplified application of the procedure to realistic cases.

\newpage
\appendix
\section{}

In Sect. 2, we have established a one-to-one correspondence between a set
of fermion states $F\equiv\{|n_1n_2\rangle\}$ (\ref{4}) and a set of boson
states $B\equiv\{|n_1n_2)\}$ (\ref{5}), both sets being orthonormal. In
correspondence with a given fermion operator ${\hat O}_F$, the mapping
procedure which we adopt searches for a boson operator ${\hat O}_B$
such that matrix elements between corresponding states be equal.
The operator ${\hat O}_B$ defines the image of ${\hat O}_F$ in $B$.

The construction of ${\hat O}_B$ proceeds step-by-step involving at each
step matrix elements between states which belong to increasingly 
larger subspaces of $F$ and $B$ \cite{samba2}. Having in mind to construct
boson images with no more than five-boson terms, it is sufficient to
involve at most the two subspaces
\[
F'=\left\{|0\rangle , \frac{1}{\sqrt{{\cal N}^F_{i}}}K_{i0}|0\rangle , 
\frac{1}{\sqrt{{\cal N}^F_{ij}}}K_{i0}K_{j0}|0\rangle ,
\frac{1}{\sqrt{{\cal N}^F_{ijk}}}K_{i0}K_{j0}K_{k0}|0\rangle \right\}
\]
and
\[
B'=\left\{|0) , b^{\dag}_i|0) , 
\frac{1}{\sqrt{{\cal N}^B_{ij}}}b^{\dag}_ib^{\dag}_{j}|0),
\frac{1}{\sqrt{{\cal N}^B_{ijk}}}b^{\dag}_ib^{\dag}_{j}b^{\dag}_{k}|0) 
\right\},
\]
where ${\cal N}^F_{i}$, ${\cal N}^F_{ij}$, 
${{\cal N}^F_{ijk}}$, ${{\cal N}^B_{ij}}$ and ${{\cal N}^B_{ijk}}$
are normalization factors.
The boson image ${\hat O}_B$ which one derives is an hermitian operator
which has the form (\ref{51}) and coefficients:
\begin{mathletters}
\begin{eqnarray}
\alpha=\langle 0|{\hat O}_F|0\rangle ,
\nonumber
\end{eqnarray}
\begin{eqnarray}
\beta_i=\frac{\langle 0|{\hat O}_FK_{i0}|\rangle}{\sqrt{{\cal N}^F_{i}}} ,
\nonumber
\end{eqnarray}
\begin{eqnarray}
\gamma_{ij}=\frac{\langle 0|K_{0i}{\hat O}_FK_{j0}|\rangle }%
{\sqrt{{\cal N}^F_{i}{\cal N}^F_{j}}}-\alpha\delta_{ij},
\nonumber
\end{eqnarray}
\begin{eqnarray}
\phi_{ij}=\frac{\langle 0|{\hat O}_FK_{i0}K_{j0}|\rangle}%
{\sqrt{{\cal N}^F_{ij}{\cal N}^B_{ij}}},
\nonumber
\end{eqnarray}
\begin{eqnarray}
\epsilon_{ijk}=
\frac{\langle 0|K_{0k}{\hat O}_FK_{i0}K_{j0}|0\rangle }%
{\sqrt{{\cal N}^F_{k}{\cal N}^F_{ij}{\cal N}^B_{ij}}}%
-\frac{\beta_i\delta_{kj}+\beta_j\delta_{ki}}{{{\cal N}^B_{ij}}},
\nonumber
\end{eqnarray}
\begin{eqnarray}
\delta_{ijkl}=%
\frac{\langle 0|K_{0i}K_{0j}{\hat O}_FK_{k0}K_{l0}|0\rangle }%
{\sqrt{{\cal N}^F_{ij}{\cal N}^F_{kl}{\cal N}^B_{ij}{\cal N}^B_{kl}}}
-\frac{\alpha\Delta^{(2)}_{ij,kl}+\sum_{i'}%
\Bigl(\gamma_{i'k}\Delta^{(2)}_{ij,i'l}+%
\gamma_{i'l}\Delta^{(2)}_{ij,i'k}\Bigr)}
{{\cal N}^B_{kl}},
\nonumber
\end{eqnarray}
\begin{eqnarray}
\rho_{ijkl}=%
\frac{\langle 0|K_{0i}K_{0j}K_{0k}{\hat O}_FK_{l0}|0\rangle }%
{\sqrt{{\cal N}^F_{ijk}{\cal N}^F_{l}{\cal N}^B_{ijk}}}
-\sum_{i'\leq j'}\phi_{i'j'}\Delta^{(3)}_{ijk,i'j'l},
\nonumber
\end{eqnarray}
\begin{eqnarray}
\tau_{ijklm}=&&
\frac{\langle 0|K_{0i}K_{0j}K_{0k}{\hat O}_FK_{l0}K_{m0}|0\rangle }%
{\sqrt{{\cal N}^F_{ijk}{\cal N}^F_{lm}{\cal N}^B_{ijk}{\cal N}^B_{lm}}}-%
\nonumber\\
&&\frac{\sum_{i'}\beta_{i'}\Delta^{(3)}_{ijk,i'lm}%
+\sum_{i'\leq j'}\epsilon_{i'j'l}\Delta^{(3)}_{ijk,i'j'm}%
+\sum_{i'\leq j'}\epsilon_{i'j'm}\Delta^{(3)}_{ijk,i'j'l}}
{{\cal N}^B_{lm}},
\nonumber
\end{eqnarray}
\end{mathletters}
where
\begin{mathletters}
\begin{eqnarray}
\Delta^{(2)}_{ij,i'j'}=(\delta_{ii'}\delta_{jj'}+%
\delta_{ij'}\delta_{ji'})/{\cal N}^B_{ij}
\nonumber
\end{eqnarray}
and
\begin{eqnarray}
\Delta^{(3)}_{ijk,i'j'k'}=&&(\delta_{ii'}\delta_{jj'}\delta_{kk'}+%
                           \delta_{ii'}\delta_{jk'}\delta_{kj'}+%
                           \delta_{ij'}\delta_{jk'}\delta_{ki'}+%
                           \delta_{ij'}\delta_{ji'}\delta_{kk'}\nonumber\\
                          &&+\delta_{ik'}\delta_{ji'}\delta_{kj'}+%
                           \delta_{ik'}\delta_{jj'}\delta_{ki'})%
/{\cal N}^B_{ijk}.
\nonumber
\end{eqnarray}
\end{mathletters}

\newpage
\begin{figure}
\caption{Ground state energy (A) and excitation energies of the lowest five
states (B) as functions of the strength $\chi /\epsilon$. Solid lines are
obtained by diagonalizing $H_F$ (\ref{3}) in $F$ (\ref{4}) while 
dot-dashed lines refer to $H_B$ (\ref{51}) in $B$ (\ref{5}).}
\label{Fig.1}
\end{figure}

\begin{figure}
\caption{
Ground state energy (A) and excitation energies of the lowest five
states (B) as functions of the strength $\chi /\epsilon$. Solid lines are
obtained by diagonalizing $H_B$ (\ref{3}) in $B$ (\ref{4}) while dot-dashed
lines are obtained with the procedure described in Sect. 2 (only 1p-1h
excitations). Numbers label different orders of approximation.
Dashed lines (B) show the RPA one-phonon energies.}
\label{Fig.2}
\end{figure}

\begin{figure}
\caption{
The same as in Fig.2 but dot-dashed lines refer now to calculations
involving up to 2p-2h excitations.}
\label{Fig.3}
\end{figure}

\begin{figure}
\caption{
Tamm-Dancoff energies as functions of the strength $\chi /\epsilon$.
~~~~~~~~~~~~~~~~~~~~~~~~~~~~~~~~~~~~~~~~~~~~~~~~}
\label{Fig.4}
\end{figure}

\begin{figure}
\caption{
Ground state energy (A) and excitation energies of the lowest five
states (B) as functions of the strength $\chi /\epsilon$. Solid lines are
obtained by diagonalizing $H_B$ (\ref{3}) in $B$ (\ref{4}) while dot-dashed
lines refer to calculations
involving only the lowest Tamm-Dancoff boson and 1p-1h excitations. 
Numbers label different orders of approximation. Further details in Sect. 5.}
\label{Fig.5}
\end{figure}

\begin{figure}
\caption{
The same as in Fig. 5 but dot-dashed lines refer now to calculations
involving up to 2p-2h excitations.}
\label{Fig.6}
\end{figure}

\newpage
\includegraphics{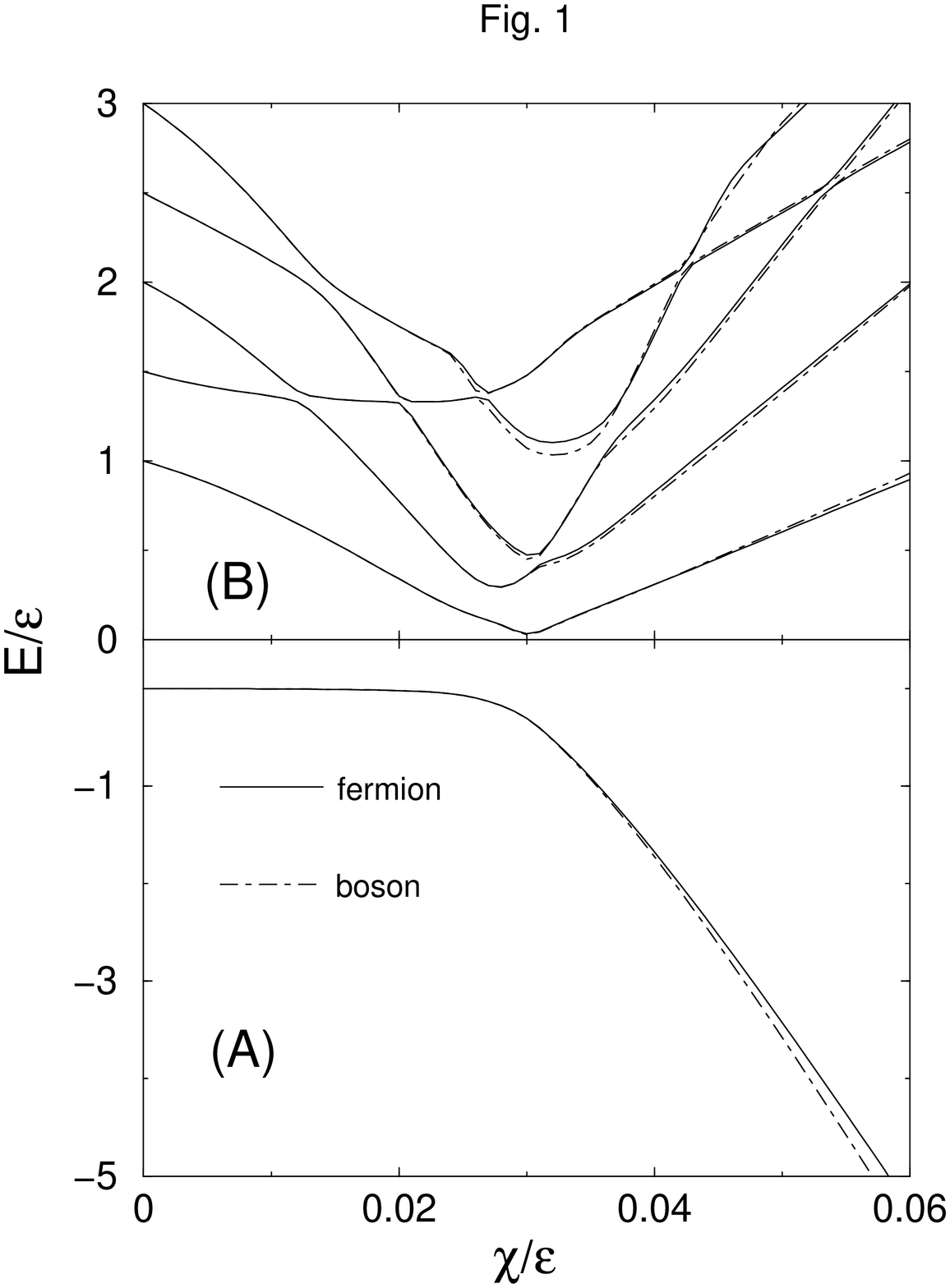}
\mbox{}\\[8cm]

\newpage
\includegraphics{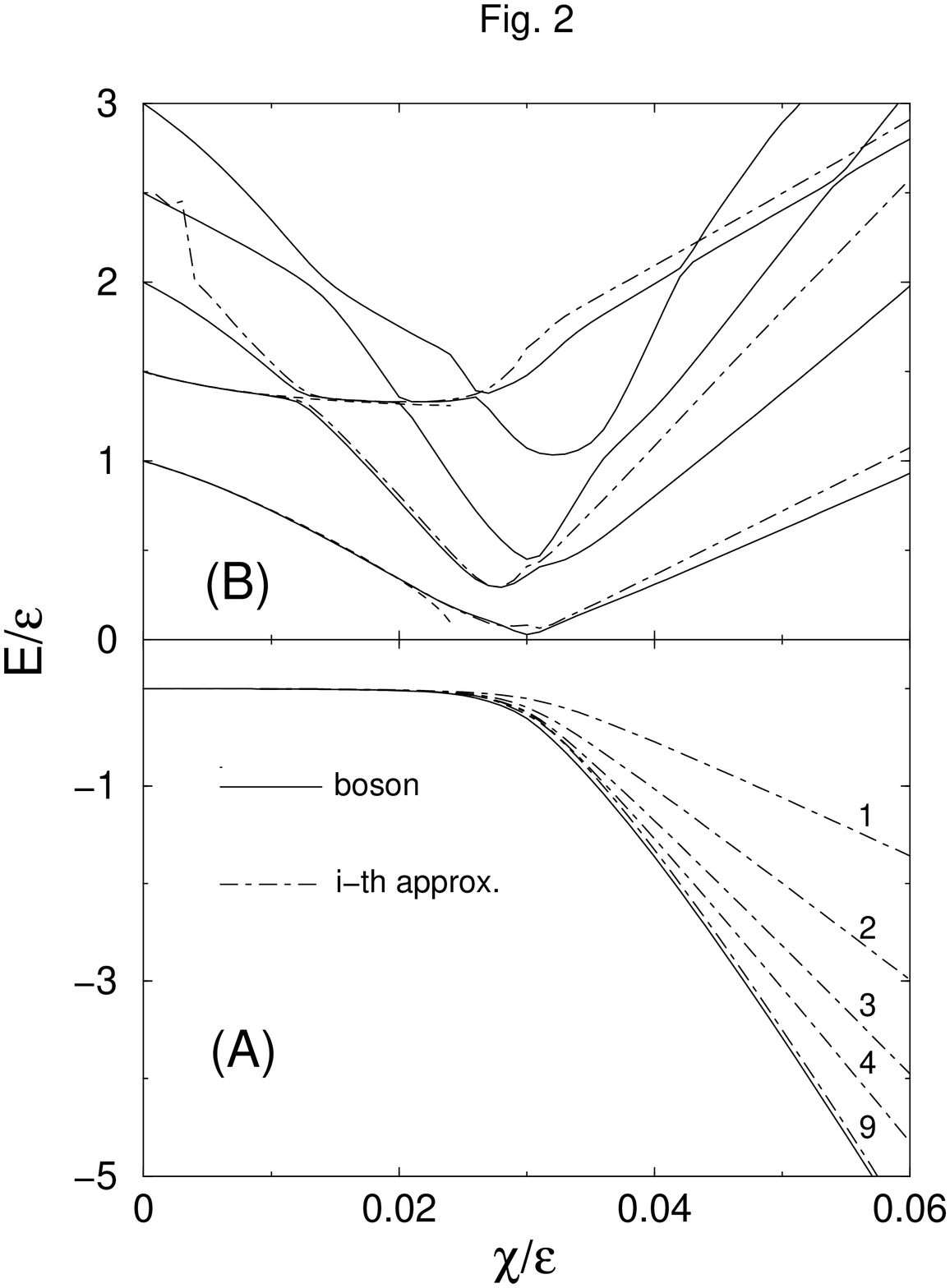}
\mbox{}\\[8cm]

\newpage
\includegraphics{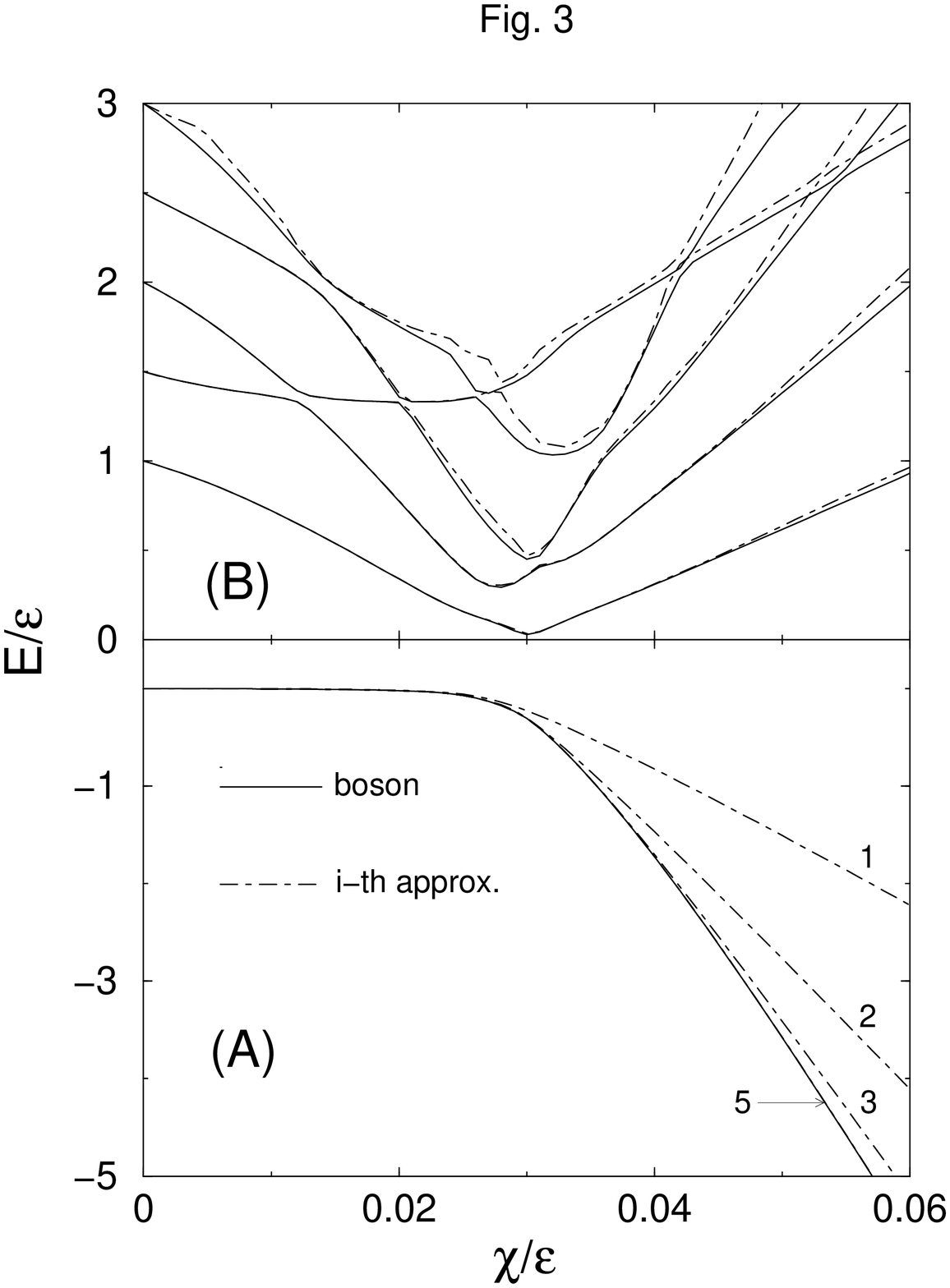}
\mbox{}\\[8cm]

\newpage
\includegraphics{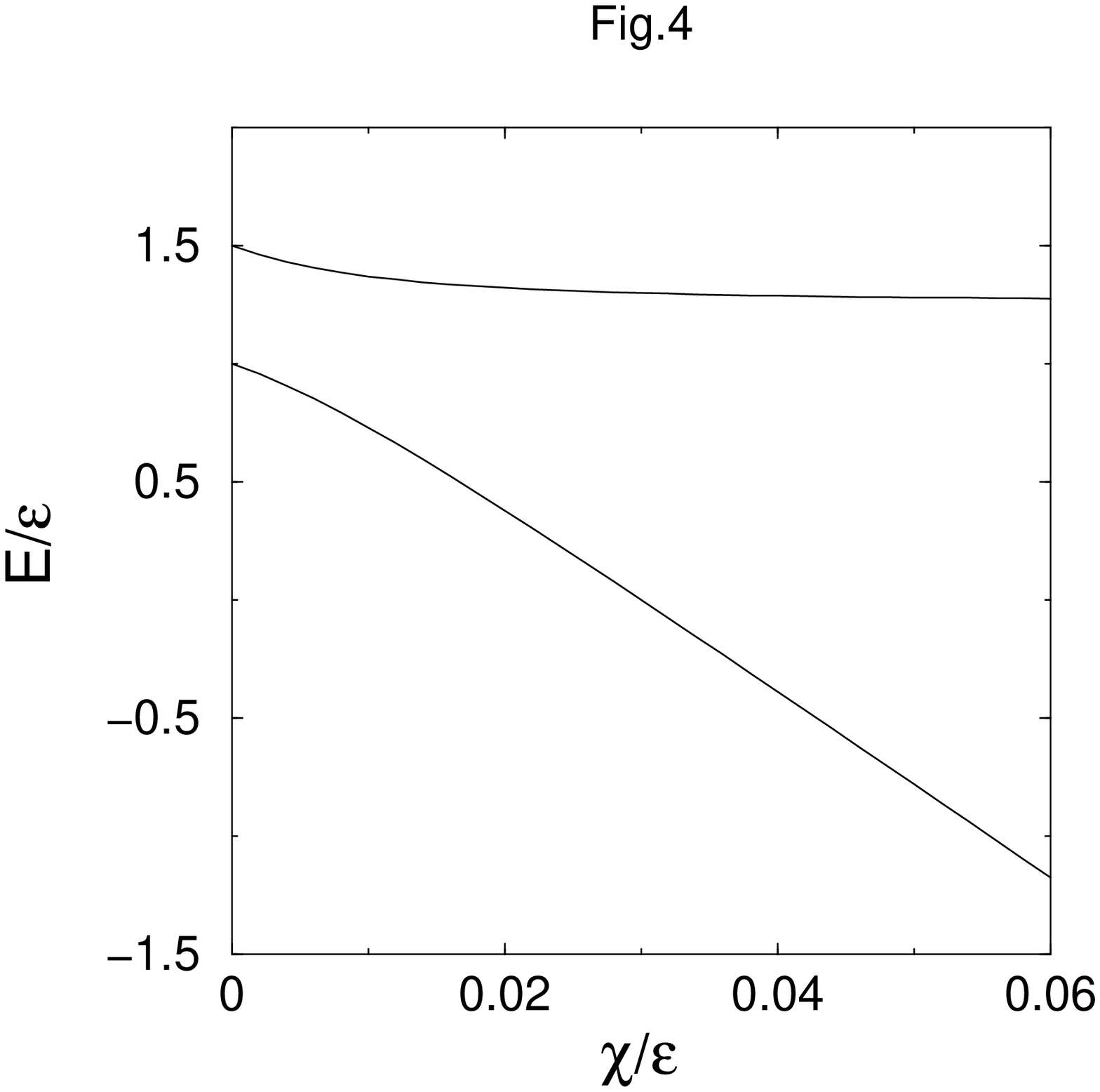}
\mbox{}\\[8cm]

\newpage
\includegraphics{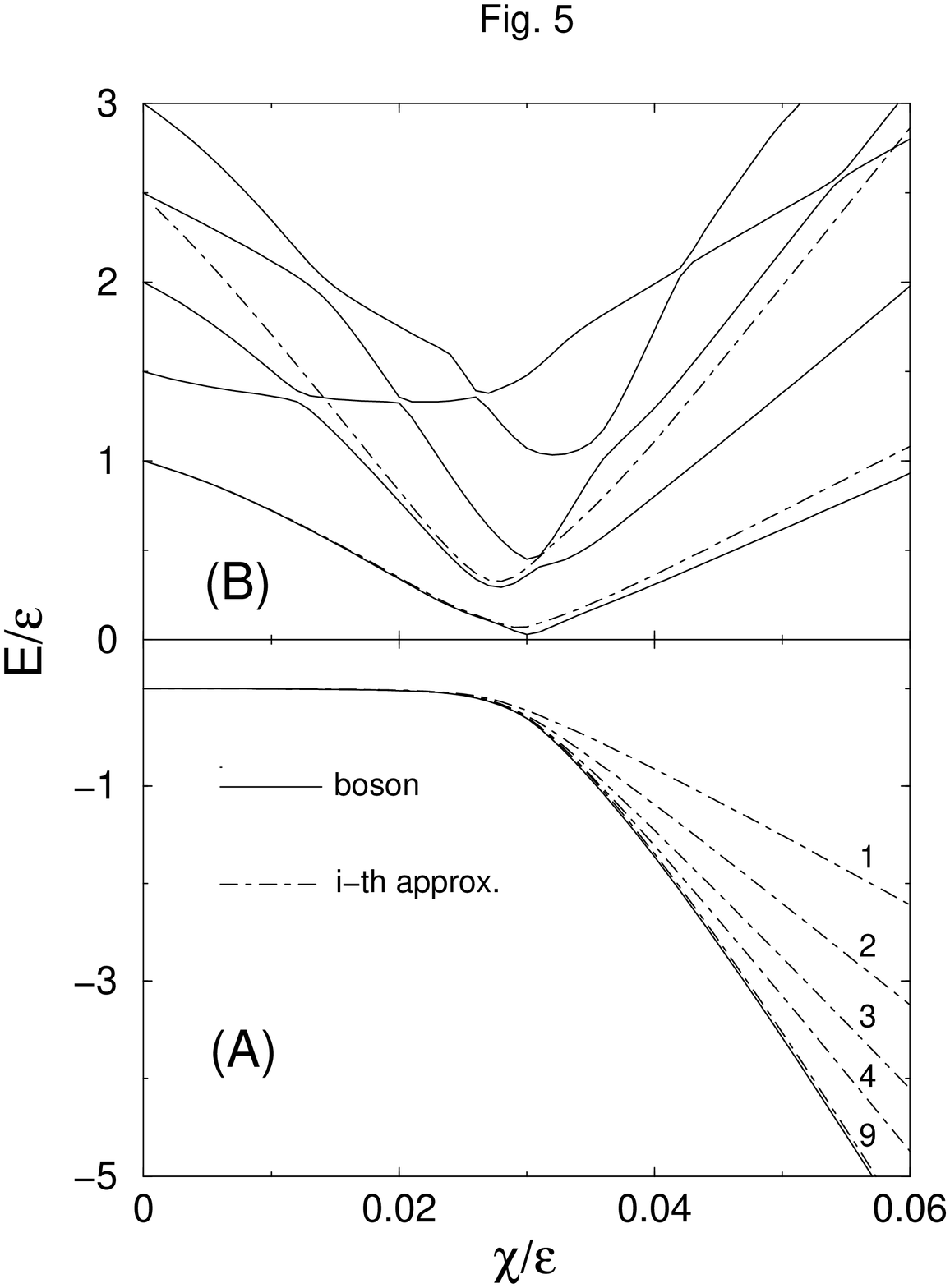}
\mbox{}\\[8cm]

\newpage
\includegraphics{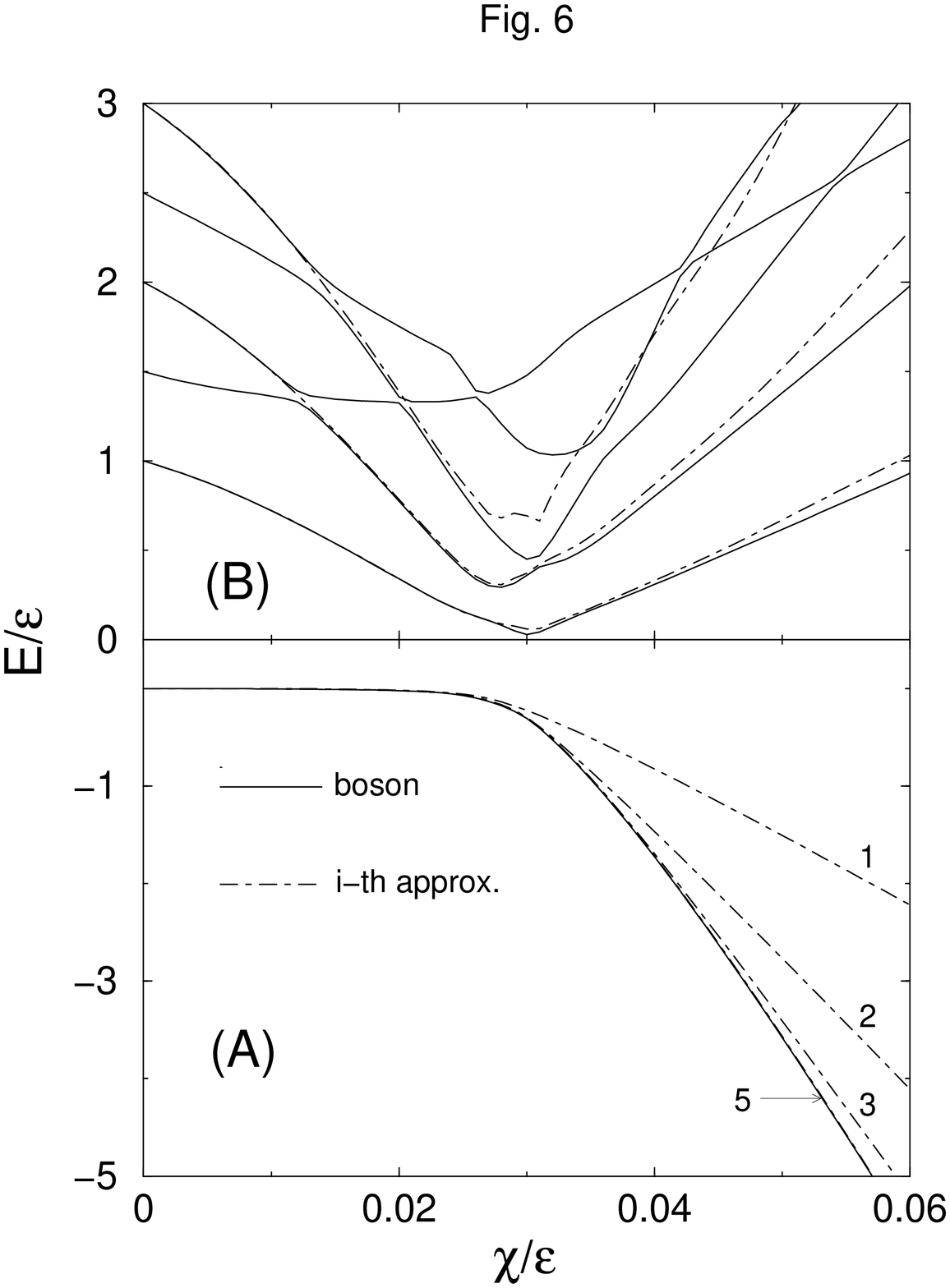}
\mbox{}\\[8cm]

\end{document}